# Hatch-Sens: a Theoretical Bio-Inspired Model to Monitor the Hatching of Plankton Culture in the Vicinity of Wireless Sensor Network

Poulami Majumder[#], Partha Pratim Ray[*]

[#] *AERU, Indian Statistical Institute,*
*203, Barrackpore Trunk Road, Kolkata, West Bengal-700108, India*
[*] *CSA Department, Sikkim University,*
*6th Mile, Samdur, PO Tadong, Gangtok, Sikkim-737102, India*

*Abstract*— Plankton research has always been an important area of biology. Due to various environmental issues and other research interests, plankton hatching and harnessing has been extremely red-marked zone for bio-aqua scientists recently. To counter this problem, no wireless sensor assisted technique or mechanism has yet not been devised. In this literature, we propose a novel approach to pursue this task by the virtue of a theoretical Bio-inspired model named Hatch-Sens, to automatically monitor different parameters of plankton hatching in laboratory environment. This literature illustrates the concepts and detailed mechanisms to accumulate this given problem. Hatch-Sens is a novel idea which combines the biology with computer in its sensing network to monitor hatching parameters of *Artemia salina*. This model reduces the manual tiresome monitoring of hatching of plankton culture by wireless sensor network.

## I. INTRODUCTION

Planktons are any organisms that live in the water column and are incapable of swimming against a current. They provide a crucial source of food to many large aquatic organisms, such as fish and whales [1].

These organisms include drifting animals, plants, archaea, or bacteria that inhabit the pelagic zone of oceans, seas, or bodies of fresh water. That is, plankton are defined by their ecological niche rather than phylogenetic or taxonomic classification [1].

Planktons [1] are primarily divided into three broad functional groups. Firstly, Phytoplankton (autotrophic, prokaryotic or eukaryotic algae) that live near the water surface where there is sufficient light to support photosynthesis. Among the more important groups are the diatoms, cyanobacteria, dinoflagellates and coccolithophores. Second, Zooplankton (protozoans or metazoans) that feed on other plankton and telonemia. Third, Bacterioplankton (bacteria and archaea), which play an important role in re-mineralizing organic material down the water column.

Plankton [1] inhabits oceans, seas, lakes, ponds. Local abundance varies horizontally, vertically and seasonally. The primary cause of this variability is the availability of light. All plankton ecosystems are driven by the input of solar energy (chemosynthesis), confining primary production to surface waters, and to geographical regions and seasons having abundant light.

In this paper, we investigate the parameters related to culture of *Artemia salina* hatching through wireless sensor network. *Artemia salina* is chosen for our purpose due to its easy availability in egged form, easy to purify and its high tolerance to a wide range of salinities. Hence, it is very helpful for lab-based work. *Artemia salina* is a species of brine shrimp – aquatic crustaceans that are more closely related to Triops and cladocerans than to true shrimp. It is very old species that have not been changed in nature for last 100 million years. Males [2] have two reproductive organs. The females can produce eggs either in the usual way or via parthenogenesis. Among two types of eggs: thin–shelled eggs that hatch immediately and thick–shelled eggs, which can remain in a dormant state. These cysts can last for a number of years, and will hatch when they are placed in water [2]. Thick–shelled eggs are produced when the body of water is drying out, raising the salt concentration. If the female dies, the eggs develop further. Eggs hatch into nauplii [6] that are about 0.5 mm in length. They have one single simple eye that only senses the presence and direction of light. Nauplii swim towards the light but adult individuals swim away from it. Later, the two more capable eyes develop but the initial eye also stays, resulting in three-eyed creatures [3]. They [2] are almost never found in an open sea, most likely because of the lack of food and relative defencelessness. However, Artemia have been observed in Elkhorn Slough, which is connected to the sea [4]. The resilience of these creatures makes them ideal test samples in experiments. *Artemia* is one of the standard organisms for testing the toxicity of chemicals [5].

We prescribe artificial hatching for our model. Artificial hatching is suitable for our need as it takes less time and monitoring headache than its natural counterpart.

Though this work is purely lab-based, we provide a conceptual model on the computational part taking the hatching job as methodical. Using wireless sensor network in such a work is unique in nature. Wireless sensor network [8] consists of spatially distributed autonomous sensors to monitor physical or environmental conditions, such as temperature, sound, vibration, pressure, humidity, motion or pollutants and to cooperatively pass their data through the network to a main location [7].The more modern networks are bi-directional, also enabling control of sensor activity. The development of wireless sensor networks was motivated by military applications such as battlefield surveillance; today such networks are used in many industrial and consumer applications, such as industrial process monitoring and control, machine health monitoring, and so on [8].

The Wireless sensor network [8] is built of "nodes" – from a few to several hundreds or even thousands, where each node is connected to one (or sometimes several) sensors. Each such sensor network node has typically several parts: a radio transceiver with an internal antenna or connection to an external antenna, a microcontroller, an electronic circuit for interfacing with the sensors and an energy source, usually a battery or an embedded form of energy harvesting. The [10] cost of sensor nodes is similarly variable, ranging from a few to hundreds of dollars, depending on the complexity of the individual sensor nodes. Size and cost constraints on sensor nodes result in corresponding constraints on resources such as energy, memory, computational speed and communications bandwidth. The topology of the WSNs can vary from a simple star network to an advanced multi-hop wireless mesh network. The propagation technique between the hops of the network can be routing or flooding [9]. WSN can be used for monitoring different applications as: area monitoring, forest fire detection, air pollution monitoring, landslide detection, machine health monitoring, industrial sense and control applications, water/wastewater monitoring, passive localization and tracking etc.

In this paper, we demonstrate the conception of Hatch-Sens to monitor the tiresome process of artificial hatching of *Artemia salina* in natural way in laboratory environment. We present a detailed view of Parmi model based on our initials: *Par*tha Pratim Ray and Poula*mi* Majumder; which is the core of Hatch-Sens. Our model monitors different hatching parameters that in turn help the fine hatching environment.

This paper is organized as follows. Section 2 presents the literature survey. Section 3 presents related work. Section 4 and 5 represents Hatch-Sens concept with Parmi as core, respectively. Section 6 is about the contribution to research. While section 7 concludes this paper.

## II. LITERATURE SURVEY

In this section we present necessary literature to understand our concept. We first provide biological studs: Plankton, hatching and culture. Then we write down computational studs: WSN and its impact on this work.

### A. Biological Studs

This section presents biological importance on these paper materials.

*1) Plankton*: Plankton [12] is a categorisation spanning a range of organism sizes including small protozoans and large metazoans. It includes holo-planktonic organisms whose complete life cycle lies within the plankton, as well as mero-planktonic organisms that spend part of their lives in the plankton before graduating to either the nekton or a sessile, benthic existence. Although Plankton is primarily transported by ambient water currents, many have locomotion, used to avoid predators (as in deil vertical migration) or to increase prey encounter rate. We perform the job on *Artemia salina* (Fig. 1).

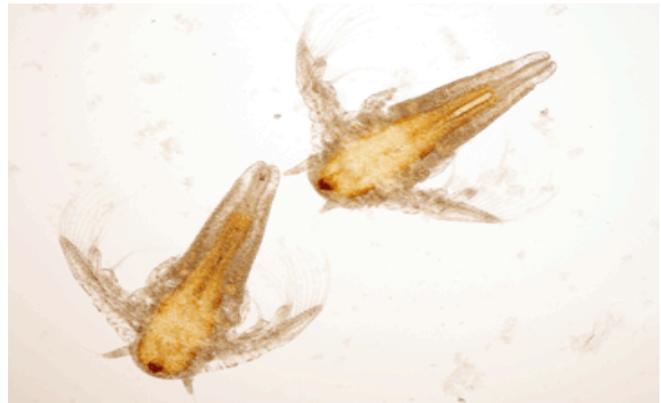

Fig. 1. *Artemia salina*

*2) Hatching:* As per Wikipedia, a hatchery [13] is a "place for artificial breeding, hatching and rearing through the early life stages of plankton, finfish and shellfish in particular". Hatcheries produce larval and juvenile aquatic livings primarily to support the aquaculture industry where they are transferred to on-growing systems i.e. fish farms to reach harvest size. Some species that are commonly raised in hatcheries include fishes, plankton etc. Additional hatchery production for small-scale laboratory uses, which is particularly prevalent in lab-based work for conservation programmes, has also yet to be quantified [14] [15]. Table 1 presents various hatching parameters.

TABLE I

*The Parameters of Hatching*

| Parameters | Description |
|---|---|
| Water | Good/Normal |
| Oxygen | Bubble $O_2$ supply |
| pH | 7.2- 8.5 |
| Illumination | Constant bright light |
| Temperature | 25 °C |
| Aeration | Needed to keep *Artemia* cysts circulating |
| Salinity | Recommended to be approximately 5-8 ppt |
| Density of cysts | Should not exceed 10 grams / liter |
| Incubation Time | Usually hatch out takes approximately 24 hours |

*3) Culture:* Culture is the method of multiplying organisms by letting them reproduce in predetermined culture media. We perform artificial culturing of *Artemia salina* in natural way. After 18-22 hours of hatching, a certain amount of yeast is added to culture media periodically, which acts as nutrient of Artemia. Few culture parameters of are as below.

- Oxygen
- pH
- Salinity
- Hardness
- Temperature
- Food

### B. Computational Studs

This portion describes wireless sensor network in brief. Wireless sensor networks consist of distributed, wirelessly enabled embedded devices capable of employing a variety of electronic sensors. Each node in a wireless sensor network is equipped with one or more sensors in addition to a microcontroller, wireless transceiver, and energy source. The microcontroller functions with the electronic sensors as well as the transceiver to form an efficient system for relaying small amounts of important data with minimal power consumption.

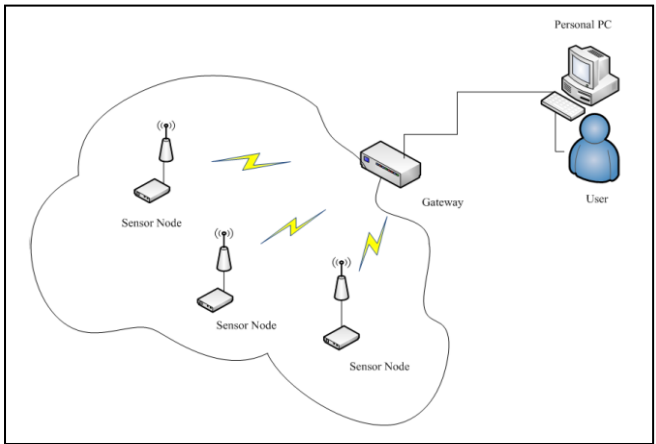

Fig. 2. Wireless sensor network

*1) Why sensor network?*

- Wireless sensor networks have many advantages over traditional sensing technology, due to their embedded construction and distributed nature. The first, and for many the most notable, feature is their cost. Using low-power and relatively inexpensive microcontrollers and transceivers, the sensor nodes used in wireless sensor networks are often less than one hundred dollars in cost.
- Another advantage wireless sensor networks hold over traditional wireless sensing technology lies in the mesh networking scheme they employ. Due to the nature of RF communication, transmitting data from one point to another using a mesh network takes less energy than transmitting directly between the two points. While embedded systems must respect their power envelope, the overall energy spent in RF communication is lower in a mesh networking scenario than using traditional point to-point communication [21].
- Sensor networks can also offer better coverage than more centralized sensing technology. Utilizing node cost advantage and mesh networking, organizations can deploy more sensors using a wireless sensor network than they could using more traditional technology [11]. This decreases the overall signal-to-noise ratio of the system, increasing the amount of usable data.

*2) Sensor node:* This network is very simple as shown in Fig. 2. User interacts with its personal PC which is preinstalled with communicating software by wire to gateway, which in turn is connected to various sensor nodes wirelessly. The main part of wireless sensor network is sensor node. A sensor node, also known as a mote, is a node in a wireless sensor network that is capable of performing some processing, gathering sensory information and communicating with other connected nodes in the network. A mote is a node but a node cannot always be a mote (Fig. 3).

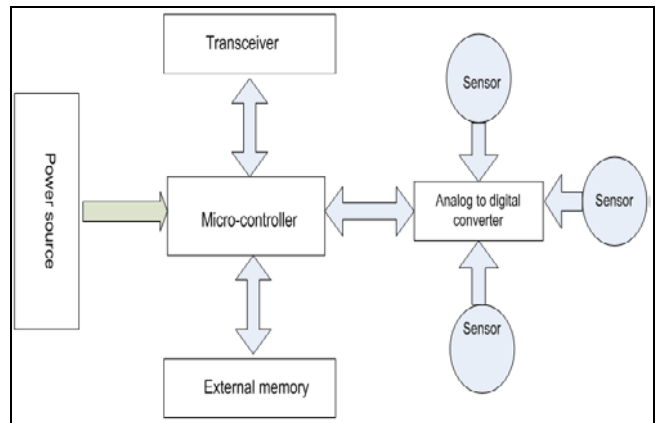

Fig. 3. Sensor node

*3) Transceiver:* Sensor nodes often make use of ISM band which gives free radio, spectrum allocation and global availability. The possible choices of wireless transmission media are Radio frequency (RF), Optical communication (Laser) Xbee and Infrared. Lasers require less energy, but need line-of-sight for communication and are sensitive to atmospheric conditions. Infrared, like lasers, needs no antenna but it is limited in its broadcasting capacity. Radio frequency based communication is the most relevant that fits most of the WSN applications. WSNs tend to use license-free communication frequencies: 173, 433, 868, and 915 MHz; and 2.4 GHz and Ethernet-based (802.11a, 802.11b, 802.11g) spread-spectrum radios. The range of license-free spread-spectrum radios depends to a large degree on the terrain, vegetation, elevation, and power levels used. The range of "out-of-the-box" Wi-Fi-Fi radios is roughly 200 m. However by adding towers, directional antennae, and amplifiers, their range can be extended to 75 kilometres (km). The most commonly used frequencies are those that can be used without a license: 900 megahertz (MHz), 2.4 gigahertz (GHz), and 5.8 GHz. These frequencies all require line-of-sight communications, and some (such as 2.4 GHz) can easily be blocked by vegetation. The functionality of both transmitter and receiver are combined into a single device known as transceivers. Transceivers often lack unique identifiers. The operational states are transmitted, receive, idle, and sleep. Current generation transceivers have built-in state machines that perform some operations automatically [17].

*4) Sensors:* A sensor (also called detector) is a converter that measures a physical quantity and converts it into a signal which can be read by an observer or by an (today mostly electronic) instrument. We use temperature sensor, humidity sensor, pH sensor, light sensor, oxygen sensor, and humidity sensor to get data about various parameters of hatching.

*5) WSN's impact:* Till date lots of research on different aspects has been performed on *Artemia salina* manually. In this paper we give a model Hatch-Sens based on WSN which reduces lot of tire work of biologist in monitoring hatching process. Parmi is core to this. Sensors take physical data and send to analog to digital converter which changes it to digital form for the understanding of micro-controller. Hence provide easy way to biologists, such a way that physical presence at laboratory is not required.

## III. RELATED WORK

*Hatch-Sens* is a novel bio-inspired model. Hence, we provide a very few literature in this paper, which have done lots of work on *Artemia salina* but for biological purposes not with any sensor based computation, for better knowledge about our task. A wildlife biologist in Michigan conducts a spring bird survey at the Green River Forestry Camp in northern New Brunswick [19] using a sophisticated system of networked acoustical sensors with over 100 acoustic sensors covering 20 listening posts in five forest types. Literature [18] review some existing uses of wireless sensor networks, identify possible areas of application, and review the underlying technologies in the hope of stimulating additional use of this promising technology to address the grand challenges of environmental science. In LUSTER [20], a fleet of sensors coordinate communications using LiteTDMA, a low-power cluster-based MAC protocol. They measure the complex light environment in thickets and are open to additional ecological parameters, such as temperature and CO2.

## IV. HATCH-SENS

*Hatch-Sens* is a bio-computation model which monitors the data sensed by sensors from various sources while hatching of *Artemia salina* in laboratory environment. Fig. 4 illustrates the idea behind hatch-sense. *Hatch-Sens* consists of two main parts. One, biological and other is computational setup. This set up is maintained at 25°C for 24 hours for monitor purpose.

### A. Biological Set up

Biological set up consists of a., b., c., d., and e. The corresponding details are as follows:
   a. Oxygen supplier: This is an electric powered machine which enables to supply oxygen to the culture through the pipes.
   b. Branched pipe: This pipe transports the $O_2$ from a. to the culture.
   c. Air pressure control keys: These keys control the vigorous supply of $O_2$ in culture clock wise.
   d. Beaker: It contains the culture of *Artemia salina*. Culture media is mixture of sea water and tap water (1:1) and net volume is 2 lit. 1 gm of *Artemia* eggs is added to this mixture. Salinity should be at 5-8 ppt (Approx.).
   e. Glass channel: It is attached inward to b. This helps the *Artemia* eggs to grow freely.

### B. Computational Set up

Computational set up consists of f., g., h., i., j., k., l., n., o., p., q., r., and s. The corresponding details are as follows:
   f. PC: Personal computer is pre installed with softwares to support the computation of Hatch-Sens.
   g. Gateway: It acts as the master device. It may be a sensor node or any other gateway.
   h. Xbee: It is the wireless communicating device run on 802.15.4 network protocol
   i. Oxygen sensor: This is analog sensor with senses the $O_2$ level of b.
   j. pH sensor: This sensor senses the pH level of the culture media.
   k. Temperature sensor: This is a digital sensor which senses the temperature.
   l. Humidity sensor: This digital sensor senses the humidity of the room.
   n. Light sensor: This is a light sensing device.
   o. Internet: Internet is used to communicate between two remote devices.
   p. User: We the human being.
   q. Smart phone: Android enabled smart phone is used to communicate with r. and s.
   r. Sensor node: This is the microcontroller board (Arduino) which act as the controlling device for the all sensors.
   s. Xbee: Same as h.

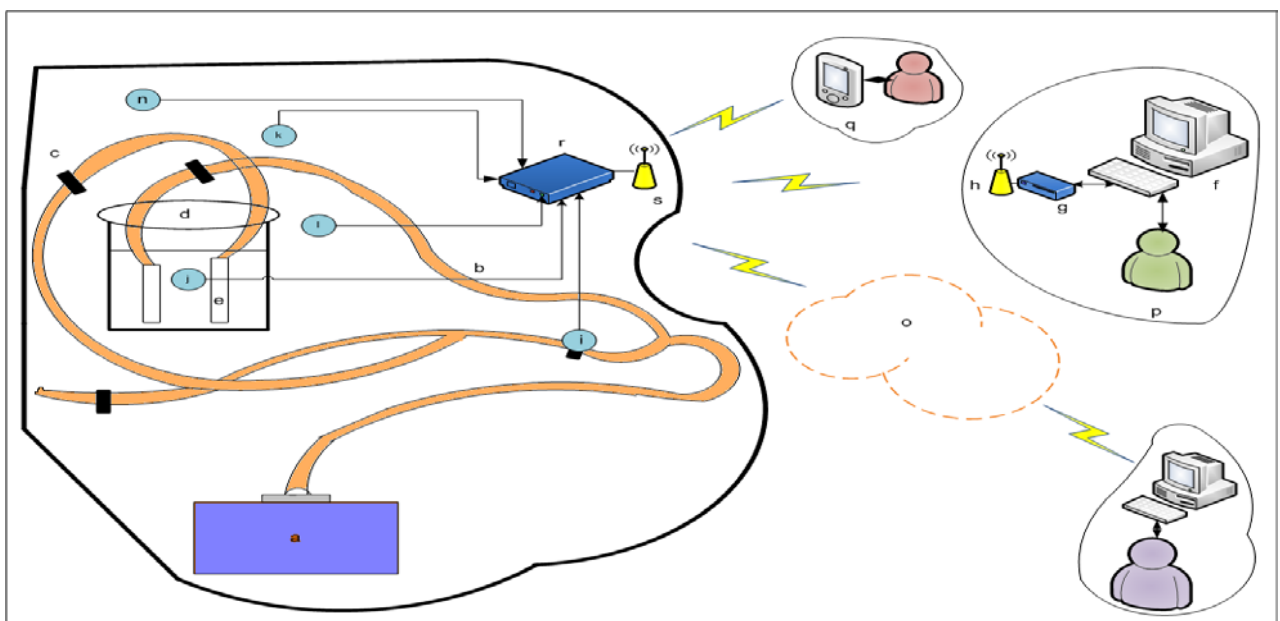

Fig. 4. Hatch-Sens

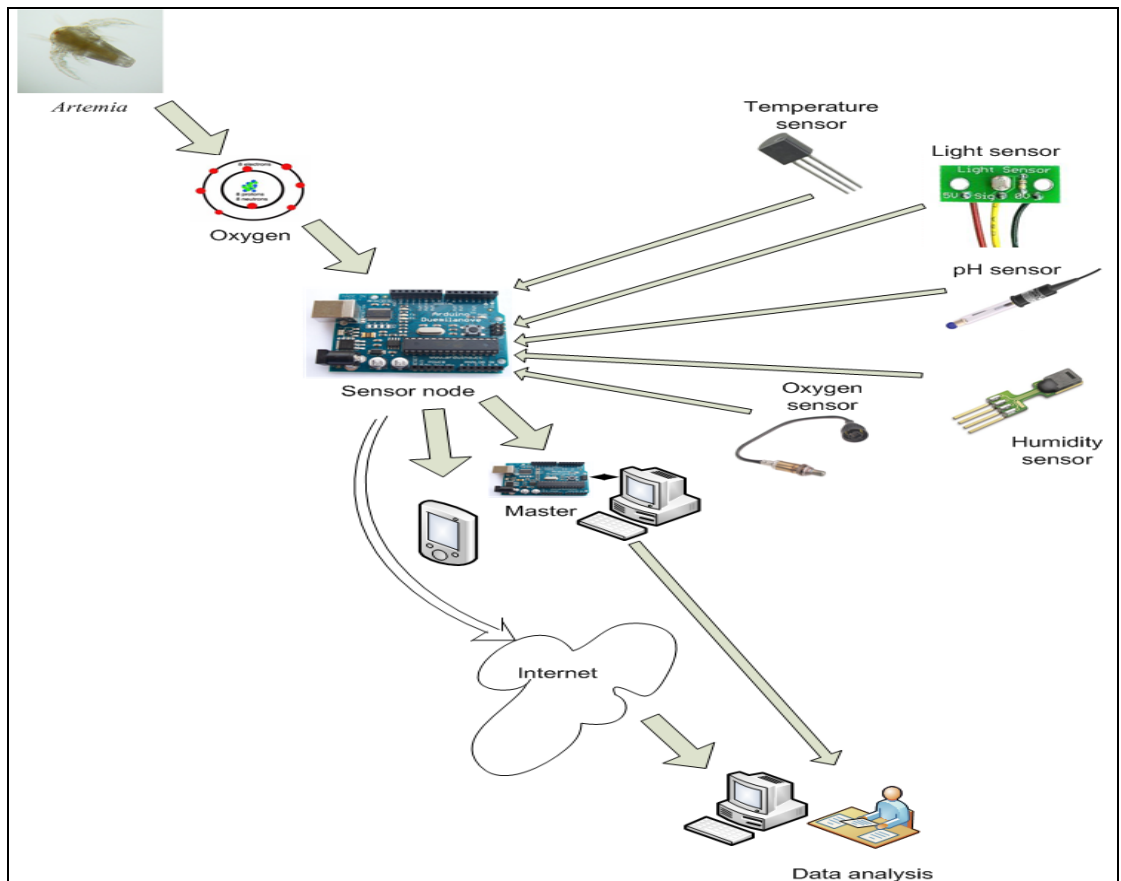

Fig. 5. Parmi mechanism

## V. PARMI

*Parmi* is a novel mechanism which describes the working of *Hatch-Sens*. This mechanism is solely responsible for the *Hatch-Sens* model to work satisfactorily. We demonstrate *Parmi* in this section. This mechanism consists of 5 steps. First: *Artemia* culture preparation, second: Oxygen supply, third: sensor node (with power) with Xbee set up, fourth: sensor connection set up, fifth: human interaction with sensor node by smart phone and/or lab-PC and/or internet, sixth: data analysis.

We describe the mechanism by Fig. 5 and the same is presented below:
 a) At first, *Artemia* culture is made ready with proper salinity and adequate media ingredients. A beaker is made up with glass pipes for oxygen blow in the culture.
 b) An oxygen supplier is powered up to blow oxygen in adequate amount into the culture, which is important for the *Artemia* to grow up.
 c) A sensor node is equipped with light, oxygen, humidity, temperature and pH sensors to get the corresponding data from the various physical media, along with an Xbee.
 d) Sensor data is collected by smart phone and/or, internet and/ or dedicated PC (a master host attached with an Xbee)

Finally the data is analysed and used for scientific (biological) work.

## VI. CONTRIBUTION

*Hatch-Sens* is a novel sensor network model which works upon *Parmi* mechanism. This work relies upon biological knowledge base about plankton specially *Artemia salina*, as we have choose it for our test purpose due to its fine biological properties. We have comprehended the problems of biologists while monitoring the hatching process of plankton in laboratory environment. Hence, we decided to design a bio-inspired model which will primarily be based on theoretical aspects of wireless sensor network. Though, this work is solely a theoretical one but it can easily be converted into its practical counterpart. So, *Hatch-Sens* performs its job to reduce the manual tiredness and work in a fantastic way. This work will surely set a new way of integrating biological (hatching of plankton) monitoring to computer network with sensors.

## VII. CONCLUSION

In this paper, we propose a novel bio-inspired model named *Hatch-Sens* to ease of constant manual monitoring process in laboratory environment. We devise *Parmi* mechanism which elaborates the practical concept behind the work. Here, we monitor important parameters such as temperature, pH, humidity, oxygen and light for betterment of *Artemia salina* hatching. We are at beginning stage of our research project which is aimed to reduce the various parameters related to monitoring of bio culture, which indeed a need of bio science community. We are heading towards the implementation of *Hatch-Sens*; taking *Parmi* as the core of it.